\documentclass[10pt,twocolumn,prl,aps,floatfix,superscriptaddress]{revtex4-1}
\usepackage{amsmath}
\usepackage{graphicx}
\usepackage{amssymb}
\def\gapx{\lower 2pt \hbox{$\buildrel>\over{\scriptstyle{\sim}}$\ }}

\newcommand{\ang}{~\text{\AA}}
\begin{document}

\title{Helium-4 Luttinger liquids in nanopores} 

\author{Adrian Del Maestro}
\affiliation{Institute for Quantum Matter and Department of Physics and Astronomy, Johns Hopkins University,
	Baltimore, MD 21218, USA}

\author{Massimo Boninsegni}
\affiliation{Department of Physics, University of Alberta, Edmonton, Alberta, Canada, T6G 2J1}

\author{Ian Affleck}
\affiliation{Department of Physics and Astronomy, University of British Columbia,
Vancouver, British Columbia V6T 1Z1, Canada}

\begin{abstract}
We study the low temperature properties of a $^4$He fluid confined in nanopores, using large-scale
Quantum Monte Carlo simulations with realistic He-He and He-pore interactions.  In the narrow-pore
limit, the system can be described by the quantum hydrodynamic theory known as Luttinger liquid
theory with a large Luttinger parameter, corresponding to the dominance of solid tendencies and
strong susceptibility to pinning by a periodic or random potential from the pore walls.  On the
other hand, for wider pores, the central region appears to behave like a Luttinger liquid with a
smaller Luttinger parameter, and may be protected from pinning by the wall potential, offering the
possibility of experimental detection of a Luttinger liquid.  
\end{abstract}

\maketitle

While the ability to flow through narrow pores is what gives a superfluid its name, a strictly
one-dimensional (1D) Galilean invariant system of bosons with short range interactions cannot exist
in any true ordered (superfluid or solid) phase. Instead, it will be in a type of quasi-ordered
phase known as a Luttinger liquid (LL), featuring correlations that decay as powers of spatial
separation, even at zero temperature ($T$) \cite{haldane}.  Ongoing experiments on $^4$He in
nanopores, have recently moved towards pore radii in the nanometer regime, thus offering the
exciting possibility of probing the LL phase.  Historically, confinement was achieved through
quasi-1D cavernous networks in porous glasses \cite{beamish} and more recently from folded sheets of
mesoporous materials \cite{taniguchi}.  A substantially different approach has been taken by Gervais
and collaborators \cite{gervais}, who are studying the flow of $^4$He inside nanopores of custom
radii by carving a hole through a Si$_3$N$_4$ membrane using a transmission electron beam.
Although the experiments of Ref.~\cite{gervais} have thus far focused on flow properties of helium
in the gas phase, it is intriguing to ponder the equilibrium system of helium atoms inside the pore
as the temperature is reduced below the bulk superfluid transition temperature $T_\lambda \simeq
2.17$~K.  If the pore radius is sufficiently small, it ought to be possible to observe a crossover
to strictly 1D behavior.

Significant progress has been made on the theoretical understanding of helium confined inside carbon
nanotubes \cite{nanotube1,nanotube2,nanotube3} or smooth nanopores \cite{nanopore1,nanopore2}.  A
complex phase diagram has been predicted, containing states where helium atoms occupy only the
central region of the cylinder (for narrow tubes) and those consisting of one or more cylindrical
shells (for wider tubes).  A realistic microscopic description of an assembly of $^4$He atoms
confined inside a single nanopore can be achieved by making use of the accepted Aziz \cite{aziz}
pair potential to describe the interaction between two helium atoms. The pore can be modeled as a
long cylindrical cavity carved inside a continuous medium. The potential energy of interaction of a
single helium atom in the pore can be obtained by integrating a Lennard-Jones pair potential, in the
same way as it is done for smooth planar substrates \cite{wall}. 

In an attempt to investigate theoretically the physical issues addressed by the planned experiments,
in this work we made use of such a model to reproduce as closely as possible the proposed
experimental geometry. An even more realistic simulation may require including the effects of a
periodic substrate, disorder in the pore walls and flow conditions, but understanding the
equilibrium state with clean smooth walls is an important first step. 

We computed low-temperature thermodynamics of fluid $^4$He inside nanopores by means of Quantum
Monte Carlo (QMC) simulations, based on the continuous-space Worm Algorithm (WA). This methodology
affords an essentially exact estimation of many physical observables, for systems of thousands of
quantum particles at low temperature, in the grand canonical ensemble \cite{worm,screw}. The
possibility of simulating large systems is crucial, as LL behavior can only be detected on
sufficiently long length scales. We chose Lennard-Jones parameters for the atom-wall potential
appropriate for Si$_3$N$_4$ \cite{siliconNitride}, as well as glass \cite{thin}, and found no
qualitative change in the basic physical results. All results shown here are for the Si$_3$N$_4$
system.

We considered nanopores of length $L$ up to $128\ang$ and radii $R$ between $2.5$ and $12.0\ang$,
using periodic boundary conditions along the axis. The illustrative results shown here are all for
$L$=100 \AA; we observed that physical results are insensitive to the length of the nanopore,
provided that $L\ \gapx 8R$. Simulations were carried out in the grand canonical ensemble, for
temperatures in the range $T = 0.5 - 2.0$~K, at a chemical potential $\mu = -7.2$~K, which
corresponds to saturated vapor pressure in the 3D reservoir used in the experiments.  In addition to
restricting $T < T_\lambda \simeq 2.17$~K, all temperatures considered are small with respect to the
average kinetic energy per particle, ensuring we are in a \emph{low-energy} regime.  A typical
simulation employs an average number of $^4$He atoms around 1000. 

%
\begin{figure}[t]
\centering
\includegraphics[width=0.8\columnwidth]{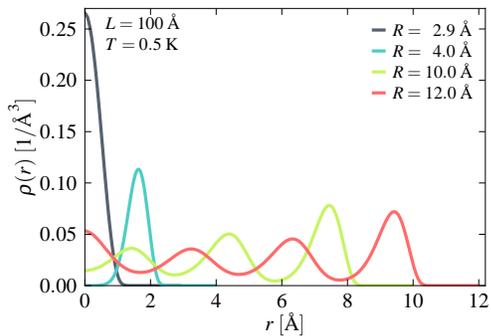}
\caption{\label{fig:radialDensity} (Color online) QMC data for the radial density of helium atoms
for cylindrical nanopores with radii $R=2.9,4.0,10.0,12.0\ang$ showing the characteristic shell
structure due to the interaction of bulk atoms with the walls of the channel. Error bars are smaller
than the line thickness.}
\end{figure}
%

Fig.~\ref{fig:radialDensity} shows the computed density $\rho(r)$ of $^4$He atoms as a function of
radial distance $r$ from the tube center, at $T = 0.5$~K.  Results are shown for
pores of different radii. The density is uniform along the axis of the nanopore ($z$), and also as
a function of angle. As expected, $\rho(r) \to 0$ in the vicinity of the wall, and features a large
peak approximately $2.5\ang$ from its surface, due to the short-distance repulsion and pronounced
minimum near the surface.  One or more peaks in the radial density are observed, corresponding to a
possible inner cylindrical region of high linear density surrounded by cylindrical shells of atoms
separated by a distance set by the attractive well of the Aziz potential.  This phenomenon is very
robust and always occurs for sufficiently large $R$.  As $R$ increases,
more peaks occur, with those near the center of the pore gradually evolving to a constant density
indicating that dimensional crossover to the 3D limit has occurred.  The presence of an inner
cylinder (IC) near the axis of the tube depends on the precise choice of $R$, and for the particular
form of the helium-pore potential used here, occurs for $R \sim 3n$ where $n\ge1$ is an integer.
Cylindrical shells in a nanopore are the analogue of planar layers in a He-4 thin film
\cite{thin,heInterface}. They arise from the suppression of quantum fluctuations near the container
wall, where atoms are localized \cite{thin}. As mentioned above, such layering has been predicted
previously in nanopores using variational and approximate Density Functional Theory
\cite{nanopore1,nanopore2}, as well as QMC at very low helium density \cite{nanopore3}.  Here, we
observe the formation of up to four such concentric shells at $\mu = -7.2$~K, owing to the relatively
large size of the systems simulated. 

An attempt to develop a theoretical understanding of the low energy properties of the helium atoms
in the pore can be made through Luttinger liquid theory \cite{haldane}, which, in strictly 1D,
provides a universal description of interacting fermions or bosons via linear quantum hydrodynamics.
This is accomplished in terms of two bosonic fields, $\theta(x)$ and $\phi(x)$, representing the
density and phase oscillations of a second quantized particle field operator $\psi^\dag(x) \sim
\sqrt{\rho_0+\partial_x \theta(x)}\mathrm{e}^{i\phi(x)}$ related to the $\mu$-dependent density of
particles at $T=0$ in the thermodynamic limit via $\rho_0 = \langle \rho(x) \rangle = \langle
\psi^\dag(x)\psi(x) \rangle$. The quadratic Hamiltonian describing these fields is given by 
\begin{equation}
H-\mu N = \frac{v}{2\pi}\int_0^L dx \left[\frac{1}{K} \left(\partial_x \phi\right)^2 + K 
\left(\partial_x \theta\right)^2\right],
\label{eq:LLHam}
\end{equation}
where we have set $\hbar=k_\text{B}=1$ and $v$ and $K$ are related to the microscopic details of the
underlying first principles many-body Hamiltonian upon which the simulations are based. The velocity
$v$ describes the linear dispersion of low energy phonon modes, while the Luttinger parameter $K$
can be tuned to initiate a $T = 0$ crossover between a state with solid order at $K = \infty$
to one with infinite range superfluid correlations at $K=0$.  The quadratic nature of
Eq.~(\ref{eq:LLHam}) allows for the calculation of all correlation functions and thermodynamic
properties in terms of $v$ and $K$.  The continuous-space WA has recently been successfully employed
\cite{llcor} to test the effectiveness of Eq.~(\ref{eq:LLHam}) in describing the low energy quantum
dynamics of a system of interacting bosons in the 1D continuum with contact interactions.  The
methodology of Ref.~\cite{llcor} is used here to study the density-density or pair correlation
function (PCF), which can be derived for a LL to be:
\begin{multline}
\langle \rho (x)\rho (0) \rangle = \rho_0^2 + \frac{1}{2\pi^2 K} \frac{d^2}{dx^2}
\ln \theta_1(\pi x/L,e^{-\pi v/LT}) \\
+ \mathcal{A} \cos (2\pi \rho_0x) \left[\frac{2\eta (iv/LT) \mathrm{e}^{-\pi v/ 6LT}}
{\theta_1(\pi x/L,e^{-\pi v/LT})}\right]^{2/K}
\label{eq:PCF}
\end{multline}
with $\eta(iz)$ the Dedekind eta function and $\theta_1(y,z)$ the first Elliptical Theta function.
$\mathcal{A}$ is a non-universal constant dependent on the short-distance properties of the system.
As $LT/v \to \infty$, Eq.~(\ref{eq:PCF}) simplifies to $\langle \rho(x) \rho(0) \rangle \to \rho_0^2
- 1/(2\pi^2 K x^2) + \mathcal{A}\cos(2\pi \rho_0 x)/x^{2/K}$ \cite{haldane}.

\begin{figure}[t]
\centering
\includegraphics[width=1\columnwidth]{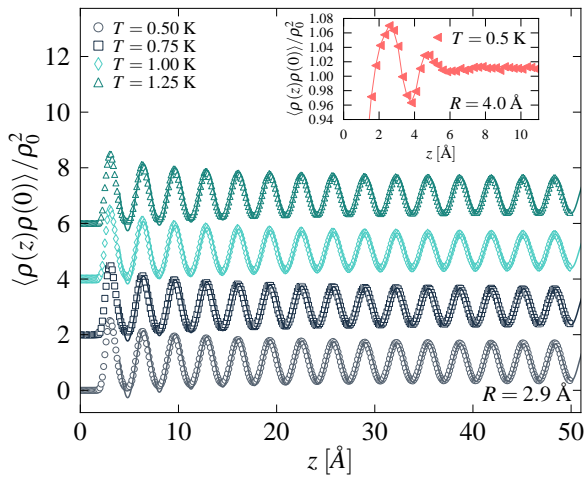}
\caption{\label{fig:pcfR2.9} (Color online) QMC data (symbols) and a fit to Eq.~(\ref{eq:PCF})
(lines) for the pair correlation function along the axis of a nanopore with $L = 100\ang$.  The main
panel shows strong oscillations for $R=2.9\ang$, while the inset details rapid decay for $R=4.0\ang$
(here the line is a guide to the eye). Error bars are smaller than the symbol size and data in the
main panel has been given a vertical $T$-dependent shift.}
\end{figure}

For $R = 2.9\ang$, Fig.~\ref{fig:radialDensity} shows that helium atoms are confined to the center
of the pore (inner cylinder) with a radial density that is effectively zero by $r \simeq 1.25\ang$;
we thus expect 1D behavior in this case.  The main panel of Fig.~\ref{fig:pcfR2.9} shows the
computed axial PCF for temperatures ranging from 0.5 -- 1.25~K. 
%
%
The results suggest that helium inside the narrow ($R=2.9\ang$) nanopore is in a quasi-solid phase, with
slowly decaying correlations only minimally dependent on temperature.  At low temperature, the
average effective 1D density ($\rho_0 = N/L$) is close to $r_\mathrm{A}^{-1}$ where $r_\mathrm{A}
\simeq 2.97\ang$ is the minimum of the helium-helium interaction potential.  In this 1D limit, we
expect the PCF to be well described by LL theory and can perform a fit of the QMC data to
Eq.~(\ref{eq:PCF}).  The He-pore interaction  is independent of the axial coordinate, and Galilean
invariance restricts the ratio $v/K = \pi \rho_0 / m$ \cite{haldane}. The finite size and
temperature scaling behavior of all thermodynamic quantities is predicted by LL theory. Thus, a
single fit of any quantity, performed for particular values of $T$ and $L$, can be used to generate
LL predictions throughout the region of $T-L$ parameter space where the universal hydrodynamic
theory of Eq.~(\ref{eq:LLHam}) applies. The results of this fitting procedure are shown as solid
lines in the main panel of Fig.~\ref{fig:pcfR2.9};  we have determined $v_{2.9} = 70(3)\ang\text{K}$
and $K_{2.9} = 6.0(2)$ where the subscript refers to the radius of the pore and the number in the
bracket is the uncertainty in the final digit. 
This large Luttinger parameter corresponds to a strong tendency toward solid formation.  The
Fig.~\ref{fig:pcfR2.9} inset shows the rapidly decaying PCF at $T=0.5$~K for $R=4.0\ang$. Here, no
IC exists, and the helium atoms form a cylindrical shell (see Fig.~\ref{fig:radialDensity}) 
with interaction energy minimizing helical density correlations at short distances that serve to
wash out possible LL oscillations at longer length scales.  We thus postpone a LL analysis of the
$R=4\ang$ pore to future studies.  In both these narrow pores, atoms experience a large degree of
localization, either along the axis or near the surface, and quantum-mechanical exchanges are
strongly suppressed.

%
\begin{figure}[t]
\centering
\includegraphics[width=1\columnwidth]{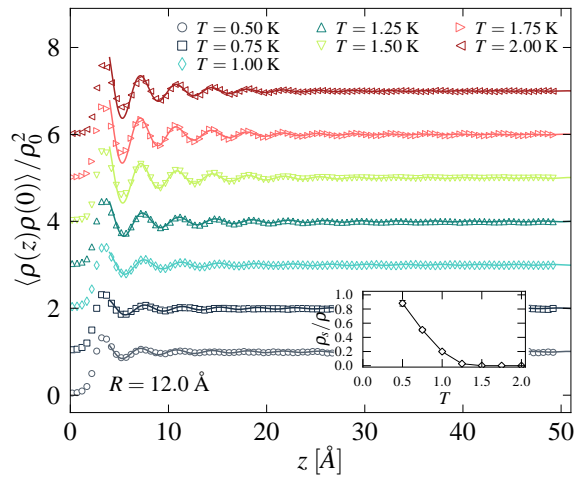}
\caption{\label{fig:pcfR12} (Color online) QMC data (symbols) and a fit to Eq.~(\ref{eq:PCF})
(lines) for the inner cylinder pair correlation function along the axis of a nanopore with $L =
100\ang$ and $R = 12.0\ang$ for helium atoms with $r < 1.75\ang$.  The inset shows the superfluid fraction
for the full nanopore as a function of temperature. Error bars are smaller than the symbol size and
data in the main panel has been given a vertical $T$-dependent shift.}
\end{figure}
%
The situation is markedly different for the $R=12.0\ang$ nanopore, as $\rho(r)$ plotted in
Fig.~\ref{fig:radialDensity} indicates the presence of an axial IC of helium, surrounded by three
cylindrical shells. 
Although the density between the shells is never strictly zero, in order to make direct comparisons
with the $R=2.9\ang$ nanopore, we can measure the properties of only those helium atoms inside the
IC defined to include all atoms with a radius smaller than the location of the first minimum in
$\rho(r)$. We find that these IC atoms make the dominant non-background (oscillatory) contribution
to the PCF and the assignment of a given indistinguishable helium atom to the IC is performed
dynamically whenever measurements are made.  QMC data for the IC-PCF are shown as symbols in
Fig.~\ref{fig:pcfR12} for temperatures ranging from 0.5 -- 2.0~K.  We observe persistent
oscillations that decay much more rapidly than in the narrower pore, with an envelope that contracts
as the temperature is reduced.  This markedly liquid-like behavior is consistent with the onset of
a finite superfluid response inside the cylinder.  The superfluid fraction can be determined through
the winding of imaginary-time particle world lines, as they wrap around the system in the axial
direction (with periodic boundary conditions) \cite{ceperleyWinding}. Results for the superfluid
density computed in this way for the full $R=12\ang$ pore ($L=100\ang$) are shown as an inset in
Fig.~\ref{fig:pcfR12} \cite{sfend}.  For comparison, the $R=2.9\ang$ nanopore of the same length
shows no evidence of superfluidity over the same temperature range. 

We may regard the $R = 12\ang$ nanopore as a coupled multi-component LL, with cylindrical shells
replacing the legs of previously studied ladders \cite{ladder}. Similar to the conclusion for a
2-leg bosonic ladder, we might expect that only one ``center of mass mode'' survives as a gapless
degree of freedom in the low energy effective field theory, due to tunneling between the shells,
manifest here as multi-particle quantum exchange cycles connecting them.  Under this assumption,
we have performed a fit of the lowest temperature QMC data in Fig.~\ref{fig:pcfR12} to
Eq.~(\ref{eq:PCF}). When considering the IC, we no longer have Galilean invariance, and must
independently extract $v$ and $K$ from the data. We find $v_{12} = 42(2)\ang\text{K}$ and $K_{12} =
1.3(1)$, which have been used to plot LL predictions (solid lines) that agree remarkably well with
the QMC data up to $T = 2.0$~K. We stress that after the LL parameter and velocity have been
determined for a single temperature, only one fit parameter $\mathcal{A}$ remains to be determined at
all other values of $T$.  The surprising robustness of the LL description of the IC can be tested
further by using the values of $v_{12}$ and $K_{12}$ determined from the PCF, to generate
predictions for other quantities.  We have compared the single-particle density matrix $\langle
\psi^\dag(x)\psi(0)\rangle$ computed in the QMC with the expected result from LL theory (with no new
fitting parameters) and find acceptable agreement at low $T$.


It is natural to ask how the form of the confining potential might affect the low energy behavior of
the nanopore system.  We have performed simulations of helium in the exact 1D limit ($R = 0$) with
$\mu = \mu_\text{3D} - V_{\text{P}}(0,2.9)$ where $V_{\text{P}}(r,R)$ is the He-pore interaction for a pore
of radius $R$. An analysis of the PCF yields values of $v_0 = 74(2)\ang\text{K}$ and $K_0 = 6.3(2)$
which are in relative agreement with those found for the $R=2.9\ang$ nanopore.  As the radius is
increased, the main consequences of the cylindrical shells of helium that surround the IC are to
screen the interaction with the wall and drastically alter the shape of the He-He potential.  The
latter effect can be quantified by computing an effective 1D interaction potential
$V_{\text{1D}}(z,R) = \frac{1}{\rho_{\text{1D}}^2} \int d^2 \zeta \int d^2 \zeta' V_\mathrm{A}
(|\vec{r} -\vec{r}'|)\rho(\zeta) \rho(\zeta')$
where $V_A$ is the Aziz pair potential, $\rho_{\text{1D}} = \int d^2 \zeta \rho(\zeta)$ the
effective 1D density, $\rho(\zeta)$ the radial number density and $\vec{r} = (\zeta,\phi,z)$ in
cylindrical polar coordinates.  We find that $V_\text{1D}(z,2.9) \approx V_\text{A}(z)$, consistent
with our expectation that this nanopore radius approximates the 1D limit.  However
$V_{\text{1D}}(z,12.0)$ exhibits a broad shallow minimum, shifted from $r_A$  with a depth that
is $1/2$ of that of the unscreened bulk potential $V_\mathrm{A}$. The softening of the effective
interaction potential decreases the energetic benefits of forming a solid with lattice spacing near
$r_A$ and as a result the helium atoms delocalize into a smaller-$K$ Luttinger liquid with a finite
superfluid density.

The value of $K_{2.9}$ found here in the 1D limit is thirty times larger that that found for a screw
dislocation of diameter $6.0\ang$ in solid $^4$He \cite{screw}. While the difference in $K$-values
may be due to the different chemical potential ($\mu =0.02$ corresponding to the bulk melting point)
or effects of the screw dislocation potential (which may vary greatly from $V_\text{P}$), it
deserves further study. 

The numerical value of the Luttinger parameter $K$ for pores of varying radius can provide important
information on the sensitivity of the LL to perturbations coming from disorder or commensuration. In
the 1D case, a weak periodic substrate, commensurate with the density is only irrelevant for $K<1/2$
\cite{haldane} while weak disorder is only irrelevant for $K<2/3$ \cite{GiamarchiSchulz}. We have
found a value of $K\approx 6$ at saturated vapor pressure for the narrowest pores, indicating
a strong tendency to form a solid, resulting from the shape of $V_A(r)$.  This large value could in
principle be experimentally tested along the lines of Ref.~\cite{gervais} as the formation of a
quasi-solid would impede the flow of helium through the pore.  Additionally, the results presented
here may be relevant to the interpretation of neutron scattering measurements of the momentum
distribution of helium in porous media \cite{glyde}, where we would expect the structure factor to
exhibit a broad feature with intensity characterized by a power law depending on $K$ at $T = 0$.  

As the radius of the pore is increased, the formation of shells near the pore wall may serve to
screen the central region from the disorder and periodic modulation of the wall potential.  Whether
this effect, plus the related reduction of the Luttinger parameter in the larger radius pores leads
to a localization length longer than the tube length, is a crucial question for the experimental
observability of LL behavior.


We thank G.~Gervais and R.~Melko for many ongoing discussions.  This work was made possible through
support from CIfAR (IA), NSERC and the U.S. Department of Energy, Office of Basic Energy Sciences,
Division of Materials Sciences and Engineering under Award DE-FG02-08ER46544 (AD).  Computational
resources were provided by WESTGRID and SHARCNET via a Compute Canada NRAC.


\begin{thebibliography}{99}
\bibitem{sfend} The observed superfluid response is a finite-size effect, i.e., it vanishes in the
$L\to\infty$ limit at fixed $T > 0$.  However, LL theory provides a prediction for the superfluid
fraction \cite{llcor} which can be used to extract further information about $v/K$.
\end{thebibliography}
\end{document}